%% file: main.tex
\documentclass{article}
\usepackage[utf8]{inputenc}
\usepackage{amsmath}
\usepackage{amssymb}
\usepackage{amsthm}
\usepackage[a-1b]{pdfx} 
\usepackage{hyperref}
\usepackage{csquotes}
\usepackage{pdflscape}
\usepackage[a4paper, total={6in, 8in}]{geometry}
\hypersetup{colorlinks,linkcolor={black},citecolor={black},urlcolor={red}}  
\usepackage[english]{babel}
\usepackage[style=apa,
    dashed=false,
	giveninits=true,
	maxcitenames=2,
	sorting =nyt,
	url=false,
	isbn =false,
	doi=false,
	eprint=false]{biblatex}
\usepackage[toc,page]{appendix}
\usepackage{comment}

\usepackage{authblk}
\providecommand{\keywords}[1]
{
  \small	
  \textbf{Keywords:} #1
}
\AtEveryBibitem{%
  \clearfield{volume}%
  \clearfield{number}}

\newcommand{\pr}{\mathbb{P}}
\newcommand{\de}{\partial}
\newcommand{\Var}{\mathbb{V}ar}
\newcommand{\Ev}{\mathbb{E}}
\newcommand{\indep}{\perp \!\!\! \perp}
\newcommand{\logitc}{\log\frac{\pr(Y=1 \mid X=x, W=w)}{\pr(Y=0 \mid X=x, W=w)}}

\newcommand{\intot}{\int_{-\infty}^{+\infty}}
\newcommand{\logitm}{\log\frac{\pr(Y=1 \mid X=x)}{\pr(Y=0 \mid X=x)}}
\DeclareMathOperator{\expit}{expit}

\usepackage{tikz}
\usetikzlibrary{arrows,positioning,shapes.geometric,decorations.markings}
\tikzstyle{squarenode}=[rectangle,draw]
\tikzstyle{littlenode}=[circle,draw,minimum size=0.8cm,font=\small]
\tikzstyle{bigellipse}=[ellipse,draw,x radius=10cm, y radius=5cm]
\tikzset{negate/.style={
            decoration={markings,
            mark= at position 0.30 with {
                  \node[yshift=13pt,transform shape] (tempnode) {$\Bigg\Vert$};
                  }
              },
              postaction={decorate}
}
}
\usepackage{subfig}

\usepackage[plain,noend]{algorithm2e}

\title{Omitting continuous covariates in binary regression models: implications for sensitivity and mediation analysis}
\author[1]{Matteo Gasparin}
\author[1]{Bruno Scarpa}
\author[2]{Elena Stanghellini}
\affil[1]{Department of Statistical Sciences, University of Padova}
\affil[2]{Department of Economics, University of Perugia}
\date{}

\begin{document}
\maketitle

\begin{abstract}
By exploiting the theory of skew-symmetric distributions, we generalise existing results in sensitivity analysis by providing the analytic expression of the bias induced by marginalization over an unobserved continuous confounder in a logistic regression model. The expression is approximated and mimics Cochran’s formula under some simplifying assumptions. Other link functions and error distributions are also considered. A simulation study is performed to assess its properties. The derivations can also be applied in causal mediation analysis, thereby enlarging the number of circumstances where simple parametric formulations can be used to evaluate causal direct and indirect effects. Standard errors of the causal effect estimators are provided via the first-order Delta method. Simulations show that our proposed estimators perform equally well as others based on numerical methods and that the additional interpretability of the explicit formulas does not compromise their precision. The new estimator has been applied to measure the effect of humidity on upper airways diseases mediated by the presence of common aeroallergens in the air.\\

\end{abstract}

\section{Introduction}

The relationship between marginal and conditional parameters in statistical models plays a key role in many investigations, ranging from sensitivity analysis to causal mediation. Under the linear least square assumption, the well-known Cochran’s formula \parencite{cochran1938} expresses the marginal parameters as a simple function of the parameters of the joint distribution. The formula generalises the so called path analysis \parencite{Wright1921} for systems of univariate linear regressions, providing a natural interpretation of the marginal parameters in terms of pathways of dependence; see \textcite{pearl2014}  or \textcite{deStavola2015mediation} for its use in causal inference. It is therefore natural to assess the extent to which it holds when the assumption of linearity is not met. 

Several papers address the issue by making use of approximated results, see the references in Section \ref{background}. \textcite{cox2007} shows that the formula does generalise, locally, to quantile regression coefficients. A condition for the formula to hold globally is also given, that however hinges on assumptions that are not suitable outside the continuous case.

The aim of this paper is to provide results for binary outcome models. We consider a simple context with a covariate of interest (called treatment) and a second variable also influencing the outcome, which may or may not be related to the treatment. We focus on a situation where the additional covariate is continuous, see \textcite{stanghellidoretti2019} for the binary case. Using the theory of skew-symmetric distributions introduced in \textcite{azzalini2013}, we provide an approximated expression of the marginal parameters that, under some simplifying assumptions, mimics the Cochran's formula thereby allowing the interpretation in terms of pathways of dependence. The derivations extend the parametric results in causal mediation analysis \parencite{pearl2001}, by providing the explicit formula of the natural effects when an interaction term between the treatment and the mediator is present. In particular, the analytic expression shows that the interaction term appears in both the numerator and the denominator, a fact generally overlooked by the existing parametric methods. The general theoretical framework here presented can be used to address other sources of nonlinearities. 

Generalisation to more complex systems of univariate regressions for both continuous and binary random variables can be made by repeatedly using the derivations here provided, opportunely  combined with  path analysis for linear equations and for binary random variables, as provided in \textcite{raggi2021} and in \textcite{lupparelli2019}. See \textcite{daniel2015causal} for the counterfactual interpretation of path-specific effects.

We start by analysing the logit link function case and then extend it to a wider class of models that includes the probit one.  The paper is structured as follows. In Section \ref{background} the theoretical background is presented together with the state of art of the literature.  The derivations are presented in Section \ref{sec:model}. Some possible extensions to the data-generating process are also proposed, followed by a simulation study where our proposal is compared with different methods exiting in the literature, including the ones commonly applied in mediation analysis, see e.g. \textcite{cheng2021estimating}. The methodology here proposed is then applied to disentangle the role of humidity and pollens on respiratory diseases based on data on urgent referrals at the Hospital of Padua between February and April 2017. 

\section{Background}
\label{background}

\begin{figure}[tb]
\centering{
\subfloat[][]{\label{fig:dag1}
\begin{tikzpicture}[scale=0.4,auto,->,>=stealth',shorten >=1pt,node distance=2.5cm] 
\node[littlenode] (W) {$W$};
\node[littlenode] (X) [below left of=W] {$X$};
\node[littlenode] (Y) [below right of=W] {$Y$};
\draw[->] (X) --node {} (W); \draw[->] (W) --node {} (Y);\draw[->] (X) --node {} (Y);
\end{tikzpicture}}\quad
\subfloat[][]{\label{fig:dag2}
\begin{tikzpicture}[scale=0.4,auto,->,>=stealth',shorten >=1pt,node distance=2.5cm] 
\node[littlenode] (W) {$W$};
\node[littlenode] (X) [below left of=W] {$X$};
\node[littlenode] (Y) [below right of=W] {$Y$};
\draw[->] (W) --node {} (Y); \draw[->] (X) --node {} (Y);
\end{tikzpicture}} \quad
\subfloat[][]{\label{fig:dag3}
\begin{tikzpicture}[scale=0.4,auto,->,>=stealth',shorten >=1pt,node distance=2.5cm] 
\node[littlenode] (W) {$W$};
\node[littlenode] (X) [below left of=W] {$X$};
\node[littlenode] (Y) [below right of=W] {$Y$};
\draw[->] (W) --node {} (Y) ; \draw[->] (W) --node {} (X); \draw[->] (X) --node {} (Y);
\end{tikzpicture}}\quad
}
\caption{Three possible data generating process of interest\label{fig:med}}
\end{figure}
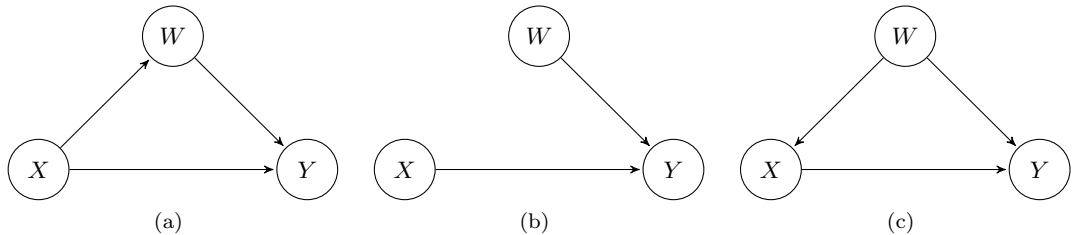

Let $X$ denote the treatment, $W$ denote the additional covariate and $Y$ denote the binary response.  Several possible data-generating processes may be of interest. We use Directed Acyclic Graphs (DAGs) to represent them, see \textcite{lauritzen1996} to which we refer for definitions. In Figure \ref{fig:dag1}, $W$ is a response of $X$ and in turn influences $Y$. In this situation, $W$ is said to be a mediator between $X$ and $Y$.  If the DAG is structural \parencite[Chapter 7]{pearl2009causality}, the decomposition of the marginal (total) effect on $Y$ of an external intervention on $X$ into a direct and indirect one, this second transmitted through $W$, is of interest.  Methods that address this issue are known as causal mediation analysis \parencite{pearl2001}. In Figure \ref{fig:dag2}, $W$ is marginally independent from $X$, a situation that may arise in controlled experiments in which $X$ is a randomised treatment and therefore $X$ and $W$ are independent by design. However $W$ is a factor that potentially influences $Y$ and understanding the link between the marginal and conditional effect of $X$ on $Y$ is therefore of scientific relevance. This investigation involves the notion of collapsibility of the effects, see \textcite{greenland2009}. In Figure \ref{fig:dag3}, $W$ is influencing both $X$ and $Y$. Sensitivity analysis accounts for the situation where $W$ is unobserved, aiming to understand how strong the association induced by the latent variable $W$ should be in order to reverse the sign of the effect of $X$ on $Y$ or, at least, to explain it away (\cite{lin1998}; \cite{vanderweele2011}).

The marginal relation between $Y$ and $X$ is given by
\begin{equation}
    \label{int_marginal}
    \pr(Y=1 \mid X=x) = \intot \pr(Y=1 \mid X=x, W=w)\, dP_{W \mid X=x}(w),
\end{equation}
and the integral can be evaluated explicitly only in favorable cases. 

As already mentioned, due to its elegance and interpretability, in the applied world several instances exist where investigators propose the use of Cochran's formula also outside the  linear case. An example is parametric mediation in causal inference for a binary outcome \parencite{vandervansteel2010} under the rare outcome assumption. Recently, the so-called {\it exact regression-based estimators} proposed in \textcite{samoilenko2022} solve the integral in \eqref{int_marginal} by using numerical integration and they prove, via simulation, that their estimators for natural effects are essentially unbiased when the model is correctly specified. However, with this method the interpretability is totally lost and no parametric formulas (exact or approximated) are developed to solve the integral. Instead, \textcite{gaynor2018} propose a closed form to the integral  based on an approximation of the inverse logit by using the cumulative density of a normal distribution. They prove in particular that their approximation is adequate when the outcome is common.  With reference to sensitivity analysis, similar derivations are \textcite{lin1998}, where, again, the analytic expression of the parameters of the marginal model is provided only for the rare outcome case. 

In order to obtain the relationship between marginal and conditional parameters, other authors such as \textcite{mackinnon2007} or \textcite*{karlson2012} start from a linear model on the latent variable underlying the binary response. The usage of a latent variable is a mathematical way to overtake the integral in \eqref{int_marginal}. In doing so, however, the mean and the variance of the dependent variable are not separately identified \parencite{winship1983} and methods have to be worked out to side step this issue. Specifically, the solution given by \textcite{karlson2012} uses the residual from the linear regression of the treatment on the mediator to compare marginal and conditional parameters of logistic regression models. This rules out the possibility of an interaction between the two regressors. In \textcite{tchetgen2014}, a closed expression for the marginal parameters is given with the imposition of a bridge distribution \parencite{wang2002} for the mediator.

We propose an analytical solution to the integral in \eqref{int_marginal} when the additional covariate $W$ is continuous. The advantages of our approach is the flexibility of the postulated data generating process that contains many commonly used model, as the linear model for $W$ and the logistic model for $Y$. We derive an approximation of the marginal parameter that closely resembles the Cochran's formula, thereby permitting a clear interpretation of the coefficient of $X$ against the log-odds ratio of $Y$ in the marginal model in terms of pathways. 

\section{Model and methods}
\label{sec:model}
We first assume the data-generating process as in Figure \ref{fig:dag1}. We then extend the derivations to cover the other two instances. In order to simplfy the notation, and without loss of generality, conditioning on a set $\bold{C}$ of covariates is omitted. We assume a logistic regression with an interaction term between the treatment and the mediator for the outcome $Y$ and a linear regression for the mediator $W$, respectively,
\begin{gather}
    \label{logit}
    \logitc = \beta_0 + \beta_x x + \beta_w w + \beta_{xw} xw,\\
    \label{lin_mod}
    \Ev[W \mid X=x] = \theta_0 + \theta_x x, \quad \varepsilon_w \sim \mathcal{N}(0, \sigma^2),
\end{gather}
The conditional distribution of the mediator given the variable $X$ is a normal with mean $\theta_0 + \theta_x x$ and variance $\sigma^2$, and its density function is given by
\begin{equation*}
    f(w \mid X=x) = \frac{1}{\sqrt{2 \pi \sigma^2}} \exp \bigg\{ -\frac{1}{2} \left( \frac{w - \theta_0 - \theta_x x}{\sigma}\right)^2 \bigg\} = \frac{1}{\sigma} \varphi \left( \frac{w - \theta_0 - \theta_x x}{\sigma}\right),
\end{equation*}
where $\varphi(\cdot)$ expresses the density function of a standard Gaussian. Then \eqref{int_marginal} becomes
\begin{equation}
    \label{int_norm}
    \begin{split}
        \pr(Y=1 \mid X=x) &= \intot \frac{\exp(\beta_0 + \beta_x x + \beta_w w + \beta_{xw}xw)}{1 + \exp(\beta_0 + \beta_x x + \beta_w w + \beta_{xw}xw)}\, d\Phi_{W\mid X=x}(w)\\
        &= \intot \frac{\exp(\beta_0 + \beta_x x + \beta_w w + \beta_{xw}xw)}{1 + \exp(\beta_0 + \beta_x x + \beta_w w + \beta_{xw}xw)} \frac{1}{\sigma} \varphi \left( \frac{w - \theta_0 - \theta_x x}{\sigma}\right)\, dw
    \end{split}
\end{equation}
where $\Phi_{W\mid X=x}(w)$ denotes the cumulative distribution function of a Gaussian distribution $\mathcal{N}(\theta_0 + \theta_x x, \sigma^2)$.

\begin{figure}
    \centering
    \includegraphics[scale=0.4]{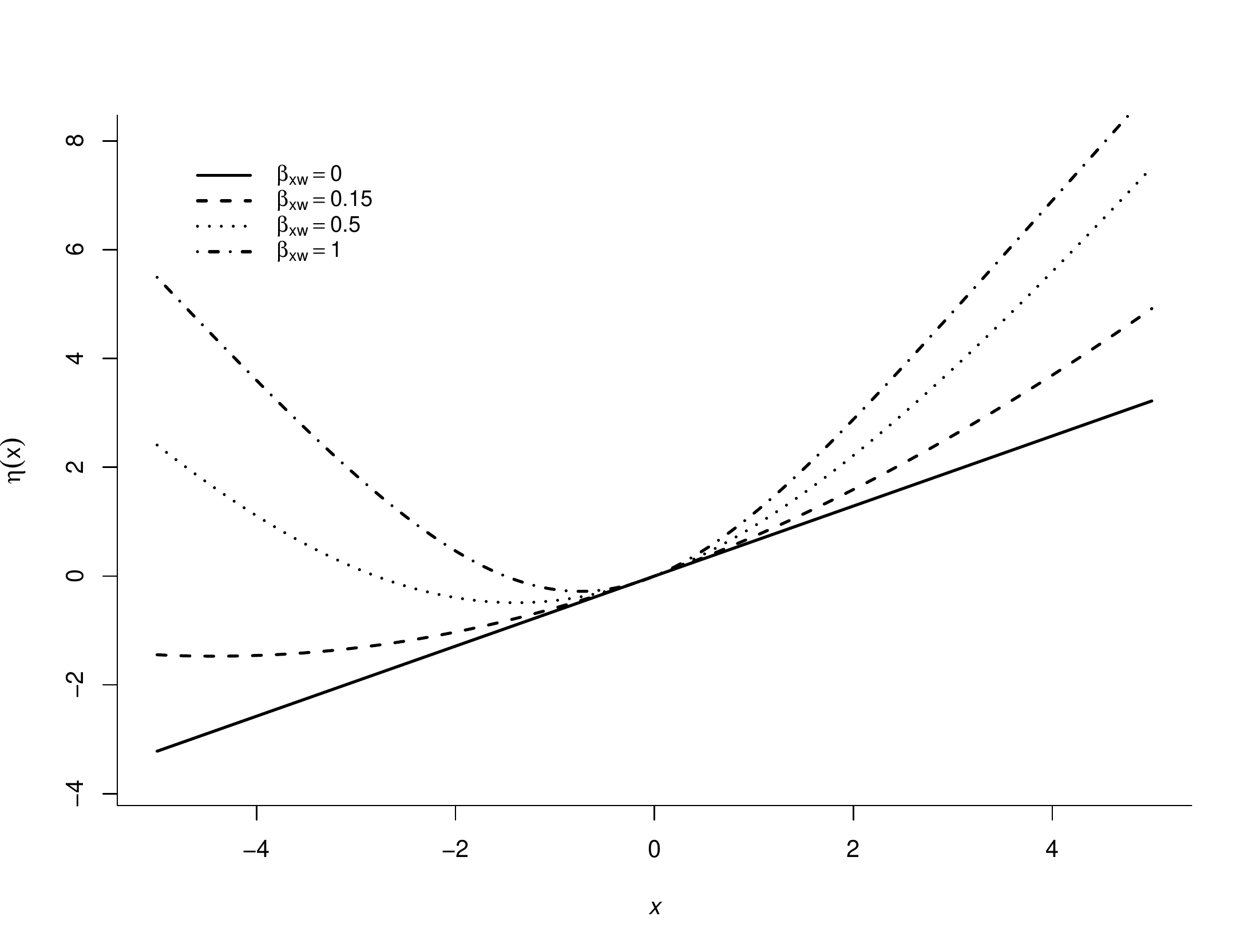}
    \caption{Function $\eta_x(x)$ with $(\theta_0, \theta_x, \sigma, \beta_x, \beta_w)=(0.1, 0.5, 0.5, 0.4, 0.5)$ and different values of $\beta_{xw}$.}
    \label{fig:eta}
\end{figure}

From standard results of skew-symmetric distributions, see \textcite{azzalini2013} p. 12-14, the integral in \eqref{int_norm} becomes
\begin{equation}
\label{exact_res_marg}
    \pr(Y=1 \mid X=x) = \pr\big\{(\beta_w + \beta_{xw}x) \sigma Z - T > - \big(\beta_0 + \beta_w\theta_0 + (\beta_x + \beta_w\theta_x + \beta_{xw}\theta_0 + \beta_{xw}\theta_x x) x\big)\big\}
\end{equation}
where $Z$ and $T$ are two independent random variables with $Z \sim \mathcal{N}(0,1)$ and $T \sim Lo(0,1)$, where $Lo(\mu,\sigma)$ indicates the Logistic distribution with $\mu$ and $\sigma$ as location and scale parameters, respectively  (see Appendix \ref{appendix_a}). It is therefore clear that the marginal logit is not linear with respect to the variable $x$ (unless in the trivial case $\beta_w = \beta_{xw} = 0$) and that the function to obtain a linear predictor in the marginal model is unknown. As a matter of fact, let $V$ be the random variable defined as
\begin{equation*}
    V = (\beta_w + \beta_{xw}x) \sigma Z - T,
\end{equation*}
it is not possible to determine analitically neither the density function nor the cumulative density function and its inverse is therefore not defined.
However,  $V$ has zero mean and it is still symmetric with bell shape and its variance is equal to
\begin{equation*}
    \Var[V] = (\beta_w + \beta_{xw}x)^2 \sigma^2 + \frac{\pi^2}{3}.
\end{equation*}
We then approximate the variable $V$ with a logistic random $V^a$. Let
\begin{equation*}
    V^a \sim Lo\bigg(0, \frac{\sqrt{3}}{\pi} \sqrt{(\beta_w + \beta_{xw}x)^2 \sigma^2 + \frac{\pi^2}{3}}\bigg),
\end{equation*}
then, using this approximation and the properties of the scale and symmetric distributions, it is possible to obtain a linear logit also for the marginal model, that is,
\begin{equation}
    \label{approx_logit}
    \begin{split}
    \logitm \approx& \log \frac{\pr\big\{V^a > -\big(\beta_0 + \beta_w\theta_0 +  (\beta_x + \beta_w\theta_x + \beta_{xw}\theta_0 + \beta_{xw}\theta_x x) x\big)\big\}}{1-\pr\big\{V^a > -\big(\beta_0 + \beta_w\theta_0 +  (\beta_x + \beta_w\theta_x + \beta_{xw}\theta_0 + \beta_{xw}\theta_x x) x\big)\big\}}\\
    =&\frac{\pi}{\sqrt{3}} \frac{\beta_0 + \beta_w\theta_0}{\sqrt{(\beta_w + \beta_{xw}x)^2 \sigma^2 + \frac{\pi^2}{3}}} + \frac{\pi}{\sqrt{3}}\frac{(\beta_x + \beta_w\theta_x + \beta_{xw}\theta_0)x + \beta_{xw}\theta_x x^2}{\sqrt{(\beta_w + \beta_{xw}x)^2 \sigma^2 + \frac{\pi^2}{3}}}.
\end{split}
\end{equation}
Equation above shows, as expected, that if there is a non-zero interaction term, the marginal model is non linear in $x$. Let $\eta_x(x)$ be 
\begin{equation}
\label{par_marginal}
\eta_x(x) = \frac{\pi}{\sqrt{3}} \frac{(\beta_x + \beta_w\theta_x + \beta_{xw}\theta_0)x + \beta_{xw}\theta_x x^2}{\sqrt{(\beta_w + \beta_{xw}x)^2 \sigma^2 + \frac{\pi^2}{3}}}.
\end{equation}
\noindent As shown in Figure \ref{fig:eta}, the function $\eta_x(x)$ mimics a quadratic function.

If, however, $\beta_{xw}=0$, then the marginal logistic regression is approximately linear with respect to the variable $x$. Let
\begin{equation}
\label{eq_marginal}
    \log \frac{\pr(Y=1 \mid X=x)}{\pr(Y=0 \mid X=x)} \approx \eta_0 + \eta_x x,
\end{equation}
\noindent be the linear approximation of the marginal logistic regression, then 
\begin{equation}
\eta_x = \frac{\pi}{\sqrt{3}} \frac{\beta_x+\beta_w\theta_x}{\sqrt{\beta_w^2 \sigma^2 + \frac{\pi^2}{3}}}
\label{eq:eta}
\end{equation}
that coincides with the Cochran's formula up to a scaling factor. This latter expression has also been given by  \textcite{mackinnon2007}. 

The situation described in Figure \ref{fig:dag2} can be addressed imposing $\theta_x=0$ in \eqref{approx_logit}, that becomes
\begin{equation*}
    \begin{split}
    \logitm \approx \frac{\pi}{\sqrt{3}} \frac{\beta_0 + \beta_w\theta_0}{\sqrt{(\beta_w + \beta_{xw}x)^2 \sigma^2 + \frac{\pi^2}{3}}} + \frac{\pi}{\sqrt{3}} \frac{\beta_x + \beta_{xw}\theta_0}{\sqrt{(\beta_w + \beta_{xw}x)^2 \sigma^2 + \frac{\pi^2}{3}}} x
    \end{split}
\end{equation*}
\noindent where $\sigma^2$ is now the marginal variance of $W$.
Some simplification arise when $\beta_{xw}=0$, as
\[
\eta_x \approx\frac{\pi}{\sqrt{3}} \frac{\beta_x}{\sqrt{\beta_w ^2 \sigma^2 + \frac{\pi^2}{3}}}. 
\]
In line with \textcite{neuhaus1993}, this result shows that, when also the interaction is null, then the marginal parameter is always smaller in modulo than the conditional parameter even when $W$ is marginally independent from $X$.  The difference between the marginal and the conditional parameter also when $\theta_x=0$ is due to the non linearity of the so-called \emph{characteristic collapsibility function} \parencite{daniel2021}, which implies the non collapsibility of the parameter in the model.

\subsection{Implications for sensitivity analysis}
To address the situation as in Figure \ref{fig:dag3}, we notice that an equivalent formulation of \eqref{int_norm}, when also $X$ is continuous, is obtained by postulating a bivariate normal for the treatment and the mediator, that is, 
\begin{equation}
    \begin{pmatrix}
    X\\
    W
    \end{pmatrix}
    \sim \mathcal{N}_2(0, \Sigma), \quad
    \Sigma = \begin{pmatrix}
        \sigma^2_x & \rho\sigma_x\\
        \rho\sigma_x & 1
    \end{pmatrix},
\end{equation}
where the parameter $\rho$ controls the correlation between the two variables. In this set-up, the regression coefficients in \eqref{lin_mod} are functions of parameters of the conditional distribution of $W$ given $X$.

In sensitivity analysis, the variable $W$ is an unmeasured factor that influences both the treatment and the outcome. Since $W$ is unmeasured, one is forced to fit the reduced model \eqref{eq_marginal}.  Such uncontrolled variable can substantially bias the estimate of the effect and lead to wrong conclusions about the relationship between the treatment and the outcome variable \parencite[Chapter 3]{vander2013}. Sensitivity analysis in the described context has been proposed by \textcite{lin1998}, but without providing the analytic expansion linking marginal and conditional parameters for the case here considered. Under the assumption of no interaction between the mediator and the treatment, we here exploit \eqref{eq:eta} to obtain an approximate relationship 
\begin{equation}
\label{rel_par}
    \beta_x \approx \eta_x \sqrt{1 + \frac{3}{\pi^2}\beta_w^2(1-\rho^2\sigma^2_x)} - \beta_w \frac{\rho}{\sigma_x},
\end{equation}
where the parameter $\sigma_x$ represents the standard deviation of the variable $X$ and it can be estimated from the data.

\begin{figure}
\centering
\includegraphics[scale=.35]{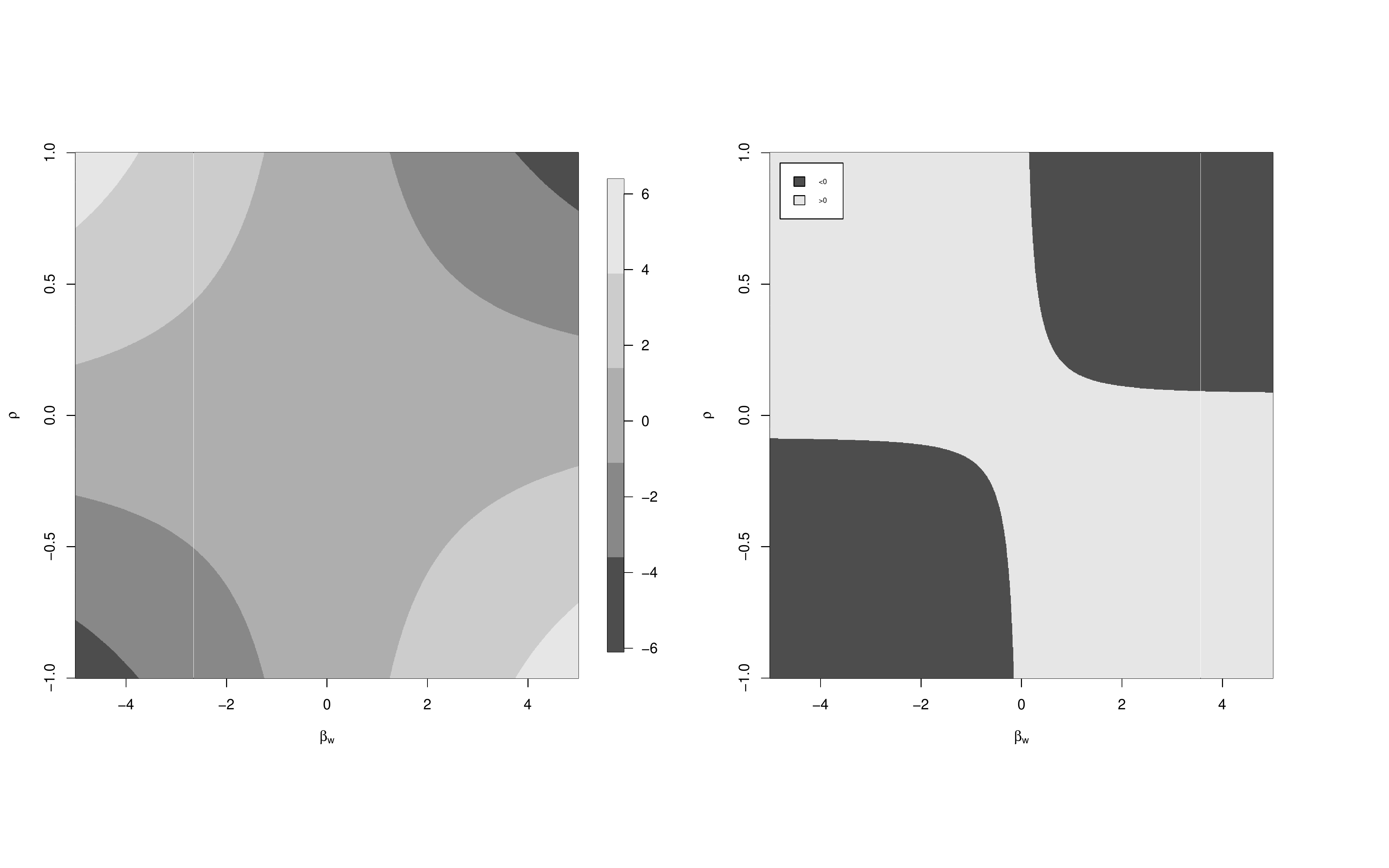}
\caption{Estimate of $\beta_x$ with different values of $\beta_w$ and $\rho$ (left) with $\eta_x=0.15$, change of sign of the ratio $\beta_x/\eta_x$ (right) with $\eta_x=0.15$.}
\label{fig:sens}
\end{figure}

A substantial simplification is achieved if also the variable $X$ is standardized. In this case, the covariance linking the variables $W$ and $X$ is simply the correlation coefficient $\rho$ and \eqref{rel_par} becomes
\begin{equation*}
    \beta_x \approx \eta_x \sqrt{1 + \frac{3}{\pi^2}\beta_w^2(1-\rho^2)} - \beta_w \rho.
\end{equation*}
Specifying plausible ranges of $\beta_w$ and $\rho$, we can obtain the effect of the treatment on the outcome simply by adjusting the marginal parameter $\eta_x$ using  \eqref{rel_par}. 
There are situations where is important to study when the sign of the conditional parameter is different from the sign of the marginal parameter. This happens when the ratio $\beta_x/\eta_x$ is smaller than 0, thus when
\[
\sqrt{1 + \frac{3}{\pi^2} \beta_w^2 (1 - \rho^2)} < \frac{\beta_w \rho}{\eta_x},
\]
an example is illustrated in Figure \ref{fig:sens}, where it is possible to see that the curve which determines the change of the sign can be approximated by a hyperbolic function.

\subsection{Extension to other link functions and error distributions}
\label{extension}
The integral in \eqref{int_marginal} can assume different forms depending on the link function chosen in \eqref{logit} and the distribution of the error in \eqref{lin_mod}. In particular, when some common conditions are verified an explicit solution of the integral is given. Let $g(\cdot)$ be a distribution function such that
\[
g(-x) = 1 - g(x),
\]
and let $\varepsilon_w$ be an error with an absolutely continuous density function $f_e(\cdot)$ defined on the real numbers with a symmetric density, so
\begin{gather*}
    f_e(x) = f_e(-x).
\end{gather*}
In some cases assumptions about zero mean and finite variance are added, although they are unnecessary in this context.

Our postulated models for the outcome given the exposure and the mediator and for the mediator given the exposure are respectively 
\begin{gather*}
    g^{-1}\{\pr(Y=1 \mid X=x, W=w)\} = \beta_0 + \beta_x x + \beta_w w + \beta_{xw} xw,\\
    W = \theta_0 + \theta_x x + \varepsilon_w,
\end{gather*}
with $\varepsilon_w$ defined as before. In this case, if the mean of $\varepsilon_w$ is well defined then the conditional mean of the mediator variable is equal to $\Ev[W \mid X=x] = \theta_0 + \theta_x x$.
So, the marginal probability of $Y$ given $X=x$ is now defined as
\begin{equation}
\label{int_general}
    \begin{split}
    \pr(Y=1 \mid X=x) =& \intot g( \beta_0 + \beta_x x + \beta_w w + \beta_{xw}xw) f_e \left(w - \theta_0 - \theta_x x\right)\, dw.\\
    =& \pr\big\{(\beta_w + \beta_{xw}x) Z - T > - \big(\beta_0 + \beta_w\theta_0 + (\beta_x + \beta_w\theta_x + \beta_{xw}\theta_0 + \beta_{xw}\theta_x x) x\big)\big\},
\end{split}
\end{equation}
where $Z$ is a random variable with density function $f_e(\cdot)$, $T \sim g$ and $Z \indep T$.

A special case appears when $\varepsilon_w \sim \mathcal{N}(0,\sigma^2)$ and $g(\cdot) = \Phi(\cdot)$, then the marginal model remains a probit regression due to the properties of the normal distribution. In addition, when $\beta_{xw}=0$, we obtain the relationship between marginal and conditional parameters in probit regression model demonstrated in \textcite{winship1983} with the use of continuous latent variables. A remarkable case is described in \textcite{tchetgen2014}, in particular if the error follows a bridge distribution \parencite{wang2002} and logit link is used for the binary regression then the marginal probability is still a standard logistic regression. Another favorable case appears to be the cauchit link function \parencite{nelder1989} and a Cauchy distribution with null location parameter for the error term $\varepsilon_w$. In this situation, the mean and the variance of the error are undefined however the convolution between the variables $Z$ and $T$ in \eqref{int_general} is closed and it is still a Cauchy. When the integral has a closed form solution, no approximations are needed and an exact formula linking the parameters of the conditional and the marginal regression is available.

\section{Causal Direct and Indirect Effects}
\label{s:Causal}
When some identifying assumptions are met, see Appendix \ref{sec:appe_caus}, the marginal effect of $X$ and $Y$ can be given a causal interpretation, as the effect of an external intervention that changes the value of the treatment from a baseline value $X=x^*$ to $X=x$ \parencite[see][]{pearl2012}. This effect is known as natural total effect $(NTE)$. Mediation analysis aims at further decomposing the total effect into a  direct and indirect effect, this second due to the mediating role of $W$. We here present the derivations underlying the decomposition of $NTE$ into natural effects given by \textcite{pearl2001}; other decompositions, see \textcite{RobinsGreenland}, can be addressed in a parallel way. 

Under further identifying assumptions, also reported in  Appendix \ref{sec:appe_caus}, the $NTE$ can be decomposed into natural direct and indirect effects via the so-called \emph{mediation formula}. Following \textcite{vandervansteel2010}, as we are dealing with a binary outcome, we here present the definition on the log odds scale. For a change in the exposure level from $X=x^*$ to $X=x$, the natural direct effect ($NDE$) and natural indirect effect ($NIE$) are
\begin{gather*}
    \log OR^{NDE}_{x,x^*} =\log \frac{h(x, x^*)}{1 - h(x, x^*)} - \log \frac{h(x^*, x^*)}{1 - h(x^*, x^*)}, \\
    \log OR^{NIE}_{x,x^*} = \log \frac{h(x, x)}{1 - h(x, x)} - \log \frac{h(x, x^*)}{1 - h(x, x^*)},
\end{gather*}
 with $\log NIE+\log NDE=\log NTE$. If our postulated data generating process is described by \eqref{logit} and \eqref{lin_mod}, then the quantity $h(x, x^*)$ is defined as
\begin{equation*}
    \label{h}
    h(x, x^*) = \intot \frac{\exp(\beta_0 + \beta_x x + \beta_w w + \beta_{xw}xw)}{1 + \exp(\beta_0 + \beta_x x + \beta_w w + \beta_{xw}xw)} \frac{1}{\sigma} \varphi \left( \frac{w - \theta_0 - \theta_x x^*}{\sigma}\right)\, dw.
\end{equation*}

Using the previous results this becomes
\begin{equation}
\label{natural_model}
\begin{split}
    h(x, x^*)&= \pr\big\{(\beta_w + \beta_{xw}x) \sigma Z - T > - (\beta_0 + \beta_w\theta_0 + (\beta_x + \beta_{xw}\theta_0 + \beta_{xw}\theta_x x) x^* + \beta_w \theta_x x^*)\big\}\\
    & \approx \expit\bigg(\frac{\pi}{\sqrt{3}} \frac{\beta_0 + \beta_w\theta_0 + (\beta_x + \beta_{xw}\theta_0 \beta_{xw}\theta_x x^*) x + \beta_w \theta_x x^*}{\sqrt{(\beta_w + \beta_{xw}x)^2 \sigma^2 + \frac{\pi^2}{3}}}\bigg)
\end{split}
\end{equation}
where the random variables $Z$ and $T$ are defined as in \eqref{exact_res_marg} while  $\expit(a) = \exp(a)/(1+\exp(a))$. Eq. \eqref{natural_model} can be used to write the natural effect model \parencite{lange2012}. 

It is therefore possible to compute the approximated closed form of $NDE$ and $NIE$ in the log-odds ratio scale, as
\begin{equation}
\label{nde}
\begin{split}
    \log OR^{NDE}_{x,x^*} \approx& \frac{\pi}{\sqrt{3}} \frac{\beta_0 + \beta_w\theta_0 + (\beta_x + \beta_{xw}\theta_0 + \beta_{xw}\theta_x x^*) x + \beta_w \theta_x x^*}{\sqrt{(\beta_w + \beta_{xw}x)^2 \sigma^2 + \frac{\pi^2}{3}}} -\\
    &+ \frac{\pi}{\sqrt{3}} \frac{\beta_0 + \beta_w\theta_0 + (\beta_x + \beta_w\theta_x + \beta_{xw}\theta_0 + \beta_{xw}\theta_x x^*)x^*}{\sqrt{(\beta_w + \beta_{xw}x^*)^2 \sigma^2 + \frac{\pi^2}{3}}},
\end{split}
\end{equation}
and 
\begin{equation}
\label{nie}
    \log OR^{NIE}_{x,x^*} \approx \frac{\pi}{\sqrt{3}} \frac{(\beta_w + \beta_{xw} x)\theta_x}{\sqrt{(\beta_w + \beta_{xw}x)^2\sigma^2 + \frac{\pi^2}{3}}} (x - x^*).
\end{equation}
This analytic expression shows that the interaction term $\beta_{xw}$ affects the natural effects in both the numerator and denominator of the expressions, a fact generally overlooked in the existing regression-based approximated methods \parencite{cheng2021estimating}.
When all effects share the same sign, the ratio $ NIE/NTE$, also known as proportion mediated, is a meaningful measure of the relative impact of the mediator on the total effect. As the denominators cancel out, the expression of the proportion mediated reduces to a more interpretable expression. 

As expected, simplifications are obtained if $\beta_{xw}=0$, as the expressions for $NDE$ and $NIE$ become
\begin{equation}
\label{no_int}
    \log  OR^{NDE}_{x,x^*} \approx \frac{\pi}{\sqrt{3}} \frac{\beta_x}{\sqrt{\beta_w^2\sigma^2 + \frac{\pi^2}{3}}}(x - x^*), \quad 
    \log  OR^{NIE}_{x,x^*} \approx \frac{\pi}{\sqrt{3}} \frac{\beta_w \theta_x}{\sqrt{\beta_w^2\sigma^2 + \frac{\pi^2}{3}}} (x - x^*).
\end{equation}
As we can notice,  expressions in \eqref{no_int} are similar to direct and indirect effects calculated through product method \parencite{baronkenny1986} in the continuous-continuous case. Estimation of the natural effects can be performed by first estimating the parameters of \eqref{logit} and \eqref{lin_mod} via standard ML methods and then plugging-in the estimated parameters in the above expressions. The assumption that $\beta_{xw}=0$ can be tested from data. In order to assess their precision, standard errors of our estimators are provided in Appendix \ref{appendix_b} and they are computed via first-order Delta method, see e.g. \textcite{casellaberger2002} pp. 240-243.

\section{Simulation study}
\label{sec:simulation}
We carried out a simulation study in order to investigate the performance of our proposed estimators for the natural effects and to compare it with the existing ones. Parameters are chosen  as in \textcite{samoilenko2022}, as follows. The variable $X$ and the mediator $W$ were simulated respectively from a $Bern(p_x)$, and a $\mathcal{N}(\theta_0 + \theta_x x, \sigma^2)$ with $p_x=0.3$, $\theta_0=0.1$, $\theta_x=0.5$ and $\sigma=0.5$. While, the response was simulated from a $Bern(p_y)$, where $\log(p_y/(1-p_y))=\beta_0 + \beta_x x + \beta_w w + \beta_{xw} xw$, with $\beta_x=0.4$, $\beta_w=0.5$, $\beta_{xw}=0.15$ and $p_y$ denoting $\pr(Y=1 \mid X=x, W=w)$. To obtain different percentages of rareness of the response, the intercept of the logistic regression is posed to $\beta_0=\{-3, -2, -0.5, 1, 2\}$ leading to the marginal prevalences, in order, of 6.65\%, 15.95\%, 44.52\%, 77.23\% and 90.03\%. In addition, every simulation is performed with 4 different sample sizes $n=\{150, 500, 1000, 5000\}$ to take into account the adequacy of the methods in different scenarios. 

The proposed methods was compared to the methods described in \textcite{vander2013} (VV), \textcite{gaynor2018} (Gaynor), \textcite{samoilenko2022} (Exact) and to the NEM approach \parencite{lange2012} (Nem). For each simulation we generated 1000 independent samples and we reported, for every combination of $\beta_0$ and $n$, the bias and the standard deviation. The true value of each scenario was computed via numerical integration based on \texttt{QUADPACK} \parencite{piessens2012}. For our method we calculated the coverage of the 95\% confidence intervals obtained via the first-order Delta method (see Appendix \ref{appendix_b}) and using percentile bootstrap, while for NEM approach the coverage of the confidence intervals was constructed using SEs based on sandwich estimator. For all other methods except NEM, 95\% confidence intervals based on percentile bootstrap with 500 replications were reported. The results for the NEM approach were obtained using the imputation approach implemented in the R package \texttt{medflex} \parencite{steen2017}. 

All the results of the simulations are shown in the Appendix \ref{sec:res_tab}. In Table \ref{tab:nde}, the performance of the various estimators of the NDE is presented. In terms of bias,  our proposed estimator performs as well as the exact and the NEM methods, and the three outperforms the others (with the exception of VV that, in the $\beta_0=-3$ and $\beta_0=-2$ scenarios, also performs well, and it is actually the best for $n=150$). Gaynor's approach performs well only when the outcome is common, in line with other simulation studies, for example, \textcite{cheng2021estimating}. Similar considerations apply to the empirical coverage of the 95$\%$ confidence interval, where, however, for small sample,  the empirical coverage of the proposed method is better 
than that of the intervals based on bootstrap (with the exception of a balanced outcome, a case in which they both perform well).  Analogous considerations apply to the performance of the estimators of the NIE, see Table \ref{tab:nie}. In line with the literature based on the continuous-continuos case, see for example \textcite{mackinnon2004confidence}, \textcite{biesanz2010assessing}, for all methods the empirical coverage of confidence intervals for NIE is good only for very large sample sizes. 

Overall, inspection of the two tables shows that the method here proposed presents satisfactory results for all configurations, unlike other methods that work well only under particular conditions of the binary response, such as rareness or commonness, as for example the approaches in \textcite{vander2013} or \textcite{gaynor2018}. In particular, it seems to offer comparable accuracy with the NEM approach and the exact regression-based method for both direct and indirect effects that, however, provide no parametric intuition of the results. A large error, due to the sparse-data bias, appears for $n=150$ and small value of the intercept in all the methods presented. A possible solution is the use of Firth's correction \parencite{firth1993} to reduce the bias of the Maximum Likelihood estimates, as proposed in \textcite{samoilenko2022}. The coverage of the CIs obtained via Delta method is adequate for all sample size bigger than 500 and for all values of $\beta_0$, and in general it is better for NDE. A proper coverage of percentile bootstrap is obtained for sample sizes bigger than 150 if our method is used.

To summarise, the precision of our proposed estimators is better than or at least comparable with the others for all levels of rareness/commonness of the outcome. In addition, bias is negligible for all $n$ and it decreases with increasing the sample size, while standard errors are the smallest in many of the proposed configurations. The additional benefit of allowing the interpretation of the effects, as functions of the parameters of the data-generating process, does not come with a loss in terms of precision.

\section{The impact of environmental factors on upper airways diseases}
 It is well-known that humidity and pollens have an impact on respiratory diseases such as rhinitis or sinusitis. Moreover, as explained in \textcite{troutt2001}, the  concentration of aeroallergens thrives with a level of humidity smaller than 50\%. Therefore, the investigation of the direct and indirect effects between environmental factors and these types of diseases is of scientific relevance.

In this Section, we used the proposed method to uncover the direct and indirect effects of  the humidity level ($X$) on upper airway (UA) diseases ($Y$), as mediated by the concentration of the  betulaceae ($W$), a very common aeroallergen. We considered 1373 observations on urgent referrals at the Hospital of Padua  between February and April 2017 \parencite{ottaviano2022}.  Humidity is represented by a dichotomic variable indicating if the daily minimum humidity in the city centre is higher than 50\%, while the concentration of betulaceae ($g/m^3$) is a continuous variable representing a daily mean metered from fixed stations and working 24h a day. Both variables refer to the day the patient was hospitalised. The binary outcome represents the presence or the absence of an upper airway (UA) disease in the patient's diagnosis. The outcome prevalence is about 49.5\%.

As we are interested in population parameters, covariates at the individual level are omitted. Parameter estimated of the selected models are presented in Table \ref{tab:res_log_reg} and \ref{tab:res_lin}.

\begin{table}
\caption{Estimates, standard errors and 90$\%$ confidence intervals of the parameters of the binary outcome model without (left panel) and  with (right panel) the treatment-mediator interaction}
\label{tab:res_log_reg}
\begin{center}
\begin{tabular}{r|rrr|rrr}
  & Estimate & Std. Error & CI 90\% & Estimate & Std. Error & CI 90\%\\ 
  \hline
  $\hat{\beta}_0$ & -1.2286 & 0.1324 & -1.4464, -1.0108 & -1.2216 & 0.1591 & -1.4834, -0.9599\\ 
  $\hat{\beta}_x$ & 0.3055 & 0.1204 & 0.1075, 0.5036 & 0.2909 & 0.2220 & -0.0743, 0.6561 \\ 
  $\hat{\beta}_w$ & 0.0087 & 0.0047 & 0.0009, 0.0164 & 0.0083 & 0.0063 & -0.0020, 0.0187\\ 
  $\hat{\beta}_{xw}$ & - & - & - & 0.0007 & 0.0095 & -0.0148, 0.0163 \\
   \hline
\end{tabular}    
\end{center}
\end{table}

\begin{table}
\begin{center}
\caption{Estimates, standard errors and 90$\%$ confidence intervals of the parameter of linear model for the mediator}
\label{tab:res_lin}
\begin{tabular}{r|rrr}
 & Estimate & Std. Error & CI 90\% \\ 
  \hline
  $\hat{\theta}_0$ & 20.4978 & 0.4842 & 19.7013, 21.2943 \\ 
  $\hat{\theta}_x$ & -2.4411 & 0.6881 & -3.5729, -1.3093 \\ 
   \hline
\end{tabular}
\end{center}
\end{table}

Table \ref{tab:res_log_reg} shows that both humidity and betualaceae have a positive effect on the outcome, with a 90\% confidence intervals suggesting the significance of the parameter estimates. The coefficients of the logistic regression when mediator-treatment is allowed are not significant at 10\% level, however both regressions allowing and not allowing interaction between $X$ and $W$ has been used to estimate casual effects as reported in Table \ref{tab:res_real}. Table \ref{tab:res_lin} shows that there is a significant negative effect of the humidity indicator on the concentration of betulaceae.

The results of our novel approach to estimate $NDE$ and $NIE$ were compared to those obtained with the methods proposed by VanderWeele and Vansteelandt \parencite{vander2013}, Gaynor \parencite{gaynor2018}, Samoilenko \parencite{samoilenko2022} and NEM method (\cite{lange2012}; \cite{steen2017}). In particular, all methods were estimated for a change in the exposure level from $X=0$ to $X=1$. For our method we computed also the estimation of the standard error via Delta method while for NEM approach robust standard errors were reported. In addition, percentile bootstrap with 2000 replications was used to calculate 90\% CIs for all methods except NEM. All methods were estimated either by allowing for the interaction between the covariates or by fixing it at zero, due to the fact that the interaction parameter in the logistic regression is mildly significant.

As we can see in Table \ref{tab:res_real}, the results  are similar to those returned by NEM approach and exact regression based method, while Gaynor and VanderWeele results are slightly higher. Confidence intervals in all proposed methods suggest the presence of a positive direct effect of humidity on the proability that an urgent referral is affected by UA desease. They also point to a negative indirect effect, as mediated through the concentration of betulaceae, due to the negative parameter $\hat{\theta}_x$ of the humidity on the mediator, which is in turn multiplied by $\hat{\beta}_w$. Together with precision, the main advantage of our method is the interpretability, indeed only VV and the proposed method offer an explicit explanation of the results as a function of the parameters, however, in the situation of a common outcome, VV tends to bias the estimate of the effects, as previously said. In general, we can easily read magnitude of the effects from the estimated parameters, a fact not so clear in the other approaches such as \textcite{samoilenko2022}.

\begin{table}
\begin{center}
\caption{Natural Direct and Indirect Effects of Humidity on the probability that a patient is affected by a UA disease as mediated by the concentration of betulaceae without (left panel) and with (right panel) interaction, different methods}
\label{tab:res_real}
\begin{tabular}{l | rrr | rrr}
  &  \multicolumn{3}{c|}{Natural Direct Effects} & \multicolumn{3}{c}{Natural Indirect Effects}\\ 
  \hline
   & Estimate & se Delta & IC 90\% boot & Estimate & se Delta & IC 90\% boot\\ 
  \hline
  Our & 0.3050 & 0.1202 & 0.1057, 0.5146 & -0.0211 & 0.0130 & -0.0451, -0.0020 \\
  VV & 0.3055 & - & 0.1058, 0.5158 & -0.0212 & - & -0.0455, -0.0021 \\
  Gaynor & 0.3171 & - & 0.1100, 0.5358 & -0.0217 & - & -0.0461, -0.0022 \\ 
  Exact & 0.3048 & - & 0.1100, 0.5358 & -0.0211 & - & -0.0450, -0.0021 \\ 
  Nem & 0.3048 & 0.1200 & - & -0.0221 & 0.0138 & - \\ 
  \hline
  Our & 0.3058 & 0.1205 & 0.1131, 0.5091 & -0.0221 & 0.0185 & -0.0445, -0.0005 \\
  VV & 0.3072 & - & 0.1131, 0.5101 & -0.0222 & - & -0.0448, -0.0005 \\
  Gaynor & 0.3181 & - & 0.1175, 0.5314 & -0.0227 & - & -0.0454, -0.0006 \\ 
  Exact & 0.3058 & - & 0.1131, 0.5086 & -0.0221 & - & -0.0443, -0.0005 \\ 
  Nem & 0.3057 & 0.1207 & - & -0.0213 & 0.0179 & - \\ 
   \hline
\end{tabular}
\end{center}
\end{table}

\section{Discussion}
Exploiting the properties of generalized-skew families \parencite{azzalini2013}, we derived the parametric expression of the integral of the marginal probability of a binary outcome when marginalisation is performed over a continuous random variable. Taking into account results coming from both sensitivity and mediation analysis, this paper contains derivations that extend existing methods in both contexts.

As for mediation, the derivations have the appealing property of providing a parametric formulation for the natural effects when no assumption is made on the rareness or commonness of the outcome and the data-generating mechanism includes an interaction between the treatment and the mediator. Simulations show that our approach performs better than or at least as well as the methods proposed in \textcite{vandervansteel2010}, \textcite{gaynor2018}, \textcite{lange2012} and \textcite{samoilenko2022}. In particular, our method performs well in all scenarios proposed and with different levels of rareness of the outcome, a featured shared only by the methods based on numerical integration. We can therefore conclude that the added value of our proposal is the interpretability of the parameters involved in the formulations, that comes with no loss in terms of precision. This interpretability is also accompanied with a low computational cost, due to the fact that the inputs of our formulations are simply the parameters of the models \eqref{logit} and \eqref{lin_mod}. Inclusion of covariates in the models can be made in a straightforward manner. When data are sparse or unbalanced, the method allows to take advantage of existing methods, such as Firth's correction. A closed form computation of the standard errors via the first-order Delta method is proposed, this formula offers an adequate quantification of the variability which is comparable with the robust standard errors used in NEM approach.   Extension to ordinal response model can be made via a cumulative logit model, as in \textcite{stanghellini2022exact} for the binary mediator. 

As noticed by \textcite{rijnhart2020}, the uptake of causal mediation for binary responses remains limited. One of the reasons is the high level of technicalities, especially when the interaction term is present. This work aims at  reconciling the results coming from linear models with the ones used in mediation analysis, providing a parametric expression of the natural effects that is as accurate as the latter but that can be easily implemented as the former. This close correspondence between the two approaches paves the way to the extension of causal inference to more complex systems, provided that the identifying assumptions are met. In these regards, as identification of causal effects crucially hinges on the assumption of no unobserved confounders, 
the sensitivity analysis here proposed can also be used to assess robustness against a continuous confounder and complement the existing cases.

\newpage
\printbibliography
\newpage

\begin{appendices}
\section{Identificability assumptions for causal effects}
\label{sec:appe_caus}
We here use the counterfactual notation as in \textcite{pearl2001} to denote with $Y_{xw}$ the random variable of $Y$ when, possibly contrary to fact, $X$ is set to $x$ and $W$ is set to $w$. This value is also called potential outcome of $Y$. Similarly, with $Y_x$ and $W_{x}$ we denote the random variable $Y$ and $W$ when $X$ is set to $x$. We further assume consistency and composition. Consistency states that the potential outcomes $Y_x$ and $W_x$ are equal to the observed ones when $X=x$ and that and so is $Y_{xw}$ when $X=x$ and $W=w$. Composition assumes that 
$Y_x = Y_{xw_{x}}$ if both $X$ and $W$ are set to their corresponding values when $X=x$.

As described in many previous works such as \textcite{vander2016}, fairly strong assumptions are needed in order to identify causal effects. Indicating by $\bold{C}$ as set of pre-exposure covariates that can be measured, the total causal effect is identified if $\bold{C}$ controls for all confouding of the treatment-outcome relation, i.e. $Y_x \indep X \mid \bold{C}$, where the symbol $\indep$ identifies the independence between the two variables. 

In addition, the set of covariates $\bold{C}$ suffices to control for confounding of both the treatment-mediator and the mediator-outcome relations, that is
\begin{gather*}
    W_x \indep X \mid \bold{C},\\
    Y_{xw} \indep X \mid \bold{C},\\
    Y_{xw} \indep W \mid X, \bold{C}.
\end{gather*}
Furthermore, in order to identify natural direct and natural indirect effects, for all levels of $x, x^*$ and $w$ the following assumption should hold true
\begin{gather*}
      Y_{xw} \indep W_{x^*} \mid \bold{C},
\end{gather*}
This latter assumption, also called \emph{cross-world independence assumption}, essentially requires that there is no confounder of  the mediator-outcome relationship that is affected by the treatment. See \textcite{steenvanstelandt} for a discussion on the identification conditions.

\section{Derivation of marginal probability}
\label{appendix_a}
The resolution of integral in Equation \eqref{int_marginal} is
\begin{equation*}
\begin{split}
    \pr(Y=1 \mid X=x) &=
     \intot \frac{\exp(\beta_0 + \beta_x x + \beta_w w + \beta_{xw}xw)}{1 + \exp(\beta_0 + \beta_x x + \beta_w w + \beta_{xw}xw)} \frac{1}{\sigma} \varphi \left( \frac{w - \theta_0 - \theta_x x}{\sigma}\right)\, dw, 
\end{split}
\end{equation*}
we apply the following variable change $s=(w - \theta_0 - \theta_x x)/\sigma$, and we obtain
\begin{equation*} 
\begin{split}
    \pr(Y=1 \mid X=x) =& \intot \expit\big(\underbrace{\beta_0 + \beta_w\theta_0 + (\beta_x + \beta_w\theta_x + \beta_{xw}\theta_0 + \beta_{xw}\theta_x x)x}_{\alpha_0}+\\
    &+ \underbrace{(\beta_w + \beta_{xw}x) \sigma}_{\alpha_s}s\big) \varphi(s) \, ds 
\end{split}
\end{equation*}
where $\expit(a) = \exp(a)/(1+\exp(a))$. In the integral appears the kernel of a skew-normal distribution described in \textcite{azzalini2013}, with normalising constant
\begin{equation*}
    \pr(Y=1 \mid X=x) = \pr\big\{(\beta_w + \beta_{xw}x) \sigma Z - T > - \big(\beta_0 + \beta_w\theta_0 + (\beta_x + \beta_w\theta_x + \beta_{xw}\theta_0 + \beta_{xw}\theta_x x) x\big)\big\}.
\end{equation*}
The integral in \eqref{h} can be solved with the similar approach.

In the general approach described in Section \ref{extension}, the integral in \eqref{int_general} can be solved as before
\begin{equation*}
\begin{split}
    \pr(Y=1 \mid X=x) =& \intot g( \beta_0 + \beta_x x + \beta_w w + \beta_{xw}xw)  f_e \left(w - \theta_0 - \theta_x x\right)\, dw, \\
    &(v.c.\,\,s=(w - \theta_0 - \theta_x x)/\sigma)\\
    =& \intot g\big(\underbrace{\beta_0 + \beta_w\theta_0 + (\beta_x + \beta_w\theta_x + \beta_{xw}\theta_0 + \beta_{xw}\theta_x x)x}_{\alpha_0} + \underbrace{(\beta_w + \beta_{xw}x)}_{\alpha_s}s\big) f_e(s) \,ds\\
    =& \pr\big\{(\beta_w + \beta_{xw}x) Z - T > - \big(\beta_0 + \beta_w\theta_0 + (\beta_x + \beta_w\theta_x + \beta_{xw}\theta_0 + \beta_{xw}\theta_x x) x\big)\big\}
\end{split}
\end{equation*}
where $Z$ is a random variable with probability density function $f_e(\cdot)$, $T \sim g$ and $Z \indep T$. 

\input{app_delta}

\input{tables}

\end{appendices}
\end{document}

%% file: app_delta.tex
\section{Delta method for standard errors of the estimators}
\label{appendix_b}
Our models are respectively
\begin{gather*}
    \logitc = \beta_0 + \beta_x x + \beta_w w + \beta_{xw} xw,\\
    W = \theta_0 + \theta_x x + \varepsilon_w, \quad \varepsilon_w \sim \mathcal{N}(0, \sigma^2),
\end{gather*}
and the parameters involved are noted by $\boldsymbol{\beta} = (\beta_0, \beta_x, \beta_w, \beta_{xw})^\top,\, \boldsymbol{\theta} = (\theta_0, \theta_x)^\top$ and $\sigma^2$ and they are estimated by using Maximum Likelihood and least squares. In particular, for $\sigma^2$ the following unbiased estimator is used
\[
\hat{\sigma}^2 = \frac{\bold{e}^\top \bold{e}}{n-p},
\]
where $p$ represents the number of parameters in the linear regression while $\bold{e}$ is the vector of residuals obtained from the linear regression. 
In addition, let $\Sigma$ be the covariance matrix of our estimators
\begin{equation}
\label{sigma}
    \Sigma = 
    \begin{pmatrix}
        \Sigma_\beta & 0 & 0\\
        0 & \Sigma_\theta & 0\\
        0 & 0 & \Sigma_{\sigma^2}\\
    \end{pmatrix},
\end{equation}
where $\Sigma_\beta, \Sigma_\theta, \Sigma_{\sigma^2}$ are respectively the covariance matrices of $\hat{\boldsymbol{\beta}}, \hat{\boldsymbol{\theta}}$ and $\hat{\sigma}^2$.
Our estimators for $NDE$ and $NIE$ proposed in Equations \eqref{nde} and \eqref{nie} are functions of these parameters. For simplicity, let us not $f_1(\boldsymbol{\beta}, \boldsymbol{\theta}, \sigma^2) = \log  OR^{NDE}_{x,x^*}$ and $f_2(\boldsymbol{\beta}, \boldsymbol{\theta}, \sigma^2) = \log  OR^{NIE}_{x,x^*}$.

\subsection{Delta method for natural direct effects}
The gradient of $f_1(\boldsymbol{\beta}, \boldsymbol{\theta}, \sigma^2)$ with respect to the parameters, denoted by 
\[
\nabla f_1(\boldsymbol{\beta}, \boldsymbol{\theta}, \sigma^2) = \bigg(\frac{\de f_1(\boldsymbol{\beta}, \boldsymbol{\theta}, \sigma^2)}{\de \boldsymbol{\beta}}, \frac{\de f_1(\boldsymbol{\beta}, \boldsymbol{\theta}, \sigma^2)}{\de \boldsymbol{\theta}}, \frac{\de f_1(\boldsymbol{\beta}, \boldsymbol{\theta}, \sigma^2)}{\de \sigma^2}\bigg)^\top,
\]
contains the following elements:
\begin{gather*}
    \frac{\de f_1(\boldsymbol{\beta}, \boldsymbol{\theta}, \sigma^2)}{\de\beta_0} = \frac{\pi}{\sqrt{3}} \Bigg\{ \frac{1}{\sqrt{(\beta_w + \beta_{xw}x)^2 \sigma^2 + \frac{\pi^2}{3}}} - \frac{1}{\sqrt{(\beta_w + \beta_{xw}x^*)^2 \sigma^2 + \frac{\pi^2}{3}}} \Bigg\}, \\
    \frac{\de f_1(\boldsymbol{\beta}, \boldsymbol{\theta}, \sigma^2)}{\de\beta_x} = \frac{\pi}{\sqrt{3}} \Bigg\{ \frac{x}{\sqrt{(\beta_w + \beta_{xw}x)^2 \sigma^2 + \frac{\pi^2}{3}}} - \frac{x^*}{\sqrt{(\beta_w + \beta_{xw}x^*)^2 \sigma^2 + \frac{\pi^2}{3}}} \Bigg\},
\end{gather*}
\begin{equation*}
\begin{split}
    &\frac{\de f_1(\boldsymbol{\beta}, \boldsymbol{\theta}, \sigma^2)}{\de\beta_w} \\
    &= \frac{\pi}{\sqrt{3}} \Bigg\{ \frac{\theta_0 + \theta_x x^*}{\sqrt{(\beta_w + \beta_{xw}x)^2 \sigma^2 + \frac{\pi^2}{3}}} - \frac{(\beta_w + \beta_{xw}x) \big(\beta_0 + \beta_w\theta_0 + (\beta_x + \beta_{xw}\theta_0 + \beta_{xw}\theta_x x^*) x + \beta_w \theta_x x^*\big) \sigma^2}{\big((\beta_w + \beta_{xw}x)^2 \sigma^2 + \frac{\pi^2}{3}\big)^{3/2}} - \\
    &\textcolor{white}{=}+ \frac{\theta_0 + \theta_x x^*}{\sqrt{(\beta_w + \beta_{xw}x^*)^2 \sigma^2 + \frac{\pi^2}{3}}} + \frac{(\beta_w + \beta_{xw}x^*) \big(\beta_0 + \beta_w\theta_0 + (\beta_x + \beta_w\theta_x + \beta_{xw}\theta_0 + \beta_{xw}\theta_x x^*) x^*\big) \sigma^2}{\big((\beta_w + \beta_{xw}x^*)^2 \sigma^2 + \frac{\pi^2}{3}\big)^{3/2}} \Bigg\},
\end{split}
\end{equation*}

\begin{equation*}
\begin{split}
    &\frac{\de f_1(\boldsymbol{\beta}, \boldsymbol{\theta}, \sigma^2)}{\de\beta_{xw}}\\
    &= \frac{\pi}{\sqrt{3}} \Bigg\{ \frac{(\theta_0 + \theta_x x^*) x}{\sqrt{(\beta_w + \beta_{xw}x)^2 \sigma^2 + \frac{\pi^2}{3}}} - \frac{(\beta_w + \beta_{xw}x) \big(\beta_0 + \beta_w\theta_0 + (\beta_x + \beta_{xw}\theta_0 + \beta_{xw}\theta_x x^*) x + \beta_w \theta_x x^*\big) x \sigma^2}{\big((\beta_w + \beta_{xw}x)^2 \sigma^2 + \frac{\pi^2}{3}\big)^{3/2}} - \\
    &\textcolor{white}{=}+ \frac{(\theta_0 + \theta_x x^*)x^*}{\sqrt{(\beta_w + \beta_{xw}x^*)^2 \sigma^2 + \frac{\pi^2}{3}}} + \frac{(\beta_w + \beta_{xw}x^*) \big(\beta_0 + \beta_w\theta_0 + (\beta_x + \beta_w\theta_x + \beta_{xw}\theta_0 + \beta_{xw}\theta_x x^*) x^*\big) x^* \sigma^2}{\big((\beta_w + \beta_{xw}x^*)^2 \sigma^2 + \frac{\pi^2}{3}\big)^{3/2}} \Bigg\},        
\end{split}
\end{equation*}
\begin{gather*}
    \frac{\de f_1(\boldsymbol{\beta}, \boldsymbol{\theta}, \sigma^2)}{\de\theta_0} = \frac{\pi}{\sqrt{3}} \Bigg\{ \frac{\beta_w + \beta_{xw} x}{\sqrt{(\beta_w + \beta_{xw}x)^2 \sigma^2 + \frac{\pi^2}{3}}} - \frac{\beta_w + \beta_{xw} x^*}{\sqrt{(\beta_w + \beta_{xw}x^*)^2 \sigma^2 + \frac{\pi^2}{3}}} \Bigg\}, \\
    \frac{\de f_1(\boldsymbol{\beta}, \boldsymbol{\theta}, \sigma^2)}{\de\theta_x} = \frac{\pi}{\sqrt{3}} \Bigg\{ \frac{(\beta_w + \beta_{xw}x) x^*}{\sqrt{(\beta_w + \beta_{xw}x)^2 \sigma^2 + \frac{\pi^2}{3}}} - \frac{(\beta_w + \beta_{xw}x^*) x^*}{\sqrt{(\beta_w + \beta_{xw}x^*)^2 \sigma^2 + \frac{\pi^2}{3}}} \Bigg\}, \\
\end{gather*}
\begin{equation*}
\begin{split}
    &\frac{\de f_1(\boldsymbol{\beta}, \boldsymbol{\theta}, \sigma^2)}{\de\sigma^2}\\
    &= \frac{\pi}{\sqrt{3}} \Bigg\{-\frac{1}{2} \frac{(\beta_w + \beta_{xw}x)^2 \big(\beta_0 + \beta_w\theta_0 + (\beta_x + \beta_{xw}\theta_0 + \beta_{xw}\theta_x x^*) x + \beta_w \theta_x x^*\big)}{\big((\beta_w + \beta_{xw}x)^2 \sigma^2 + \frac{\pi^2}{3}\big)^{3/2}} + \\
    &\textcolor{white}{=}+ \frac{1}{2} \frac{(\beta_w + \beta_{xw}x^*)^2 \big(\beta_0 + \beta_w\theta_0 + (\beta_x + \beta_w\theta_x + \beta_{xw}\theta_0 + \beta_{xw}\theta_x x^*) x^*\big)}{\big((\beta_w + \beta_{xw}x^*)^2 \sigma^2 + \frac{\pi^2}{3}\big)^{3/2}} \Bigg\},
\end{split}
\end{equation*}

So, the standard error for the estimator of natural direct effects is equal to 
\begin{equation*}
    se(\widehat{\log OR^{NDE}_{x,x^*}}) = \sqrt{\nabla f_1(\hat{\boldsymbol{\beta}}, \hat{\boldsymbol{\theta}}, \hat{\sigma}^2)^\top\, \Sigma \, \nabla f_1(\hat{\boldsymbol{\beta}}, \hat{\boldsymbol{\theta}}, \hat{\sigma}^2)},
\end{equation*}
where $\Sigma$ is defined in \eqref{sigma}. An huge simplification is obtained if $\beta_{xw}$ is fixed at 0; in this case $\boldsymbol{\beta}=(\beta_0, \beta_x, \beta_w)$, and the elements $\de f_1(\boldsymbol{\beta}, \boldsymbol{\theta}, \sigma^2)/ \de\beta_0, \, \de f_1(\boldsymbol{\beta}, \boldsymbol{\theta}, \sigma^2)/ \de\theta_0$ and $\de f_1(\boldsymbol{\beta}, \boldsymbol{\theta}, \sigma^2)/ \de\theta_x$ are constantly equal to zero.

\subsection{Delta method for natural indirect effects}
The gradient of $f_2(\boldsymbol{\beta}, \boldsymbol{\theta}, \sigma^2)$ with respect to the parameters, denoted by 
\[
\nabla f_2(\boldsymbol{\beta}, \boldsymbol{\theta}, \sigma^2) = \bigg(\frac{\de f_2(\boldsymbol{\beta}, \boldsymbol{\theta}, \sigma^2)}{\de \boldsymbol{\beta}}, \frac{\de f_1(\boldsymbol{\beta}, \boldsymbol{\theta}, \sigma^2)}{\de \boldsymbol{\theta}}, \frac{\de f_1(\boldsymbol{\beta}, \boldsymbol{\theta}, \sigma^2)}{\de \sigma^2}\bigg)^\top,
\]
contains the following elements:
\begin{gather*}
    \frac{\de f_2(\boldsymbol{\beta}, \boldsymbol{\theta}, \sigma^2)}{\de \beta_0} = 0, \quad
    \frac{\de f_2(\boldsymbol{\beta}, \boldsymbol{\theta}, \sigma^2)}{\de \beta_x} = 0\\
    \frac{\de f_2(\boldsymbol{\beta}, \boldsymbol{\theta}, \sigma^2)}{\de \beta_w} = \frac{\pi}{\sqrt{3}}(x - x^*) \Bigg\{\frac{\theta_x}{\sqrt{(\beta_w + \beta_{xw}x)^2 \sigma^2 + \frac{\pi^2}{3}}} - \frac{(\beta_w + \beta_{xw}x)^2 \sigma^2}{\big((\beta_w + \beta_{xw}x)^2 \sigma^2 + \frac{\pi^2}{3}\big)^{3/2}}\Bigg\}\\
    \frac{\de f_2(\boldsymbol{\beta}, \boldsymbol{\theta}, \sigma^2)}{\de \beta_{xw}} = \frac{\pi}{\sqrt{3}}(x - x^*) \Bigg\{\frac{\theta_x x}{\sqrt{(\beta_w + \beta_{xw}x)^2 \sigma^2 + \frac{\pi^2}{3}}} - \frac{(\beta_w + \beta_{xw}x)^2 x \sigma^2}{\big((\beta_w + \beta_{xw}x)^2 \sigma^2 + \frac{\pi^2}{3}\big)^{3/2}}\Bigg\}\\
    \frac{\de f_2(\boldsymbol{\beta}, \boldsymbol{\theta}, \sigma^2)}{\de \theta_0} = 0\\
    \frac{\de f_2(\boldsymbol{\beta}, \boldsymbol{\theta}, \sigma^2)}{\de \theta_x} = \frac{\pi}{\sqrt{3}}(x - x^*) \Bigg\{\frac{\beta_w + \beta_{xw}x}{\sqrt{(\beta_w + \beta_{xw}x)^2 \sigma^2 + \frac{\pi^2}{3}}}\Bigg\}\\
    \frac{\de f_2(\boldsymbol{\beta}, \boldsymbol{\theta}, \sigma^2)}{\de \sigma^2} = \frac{\pi}{\sqrt{3}}(x - x^*) \Bigg\{-\frac{1}{2} \frac{(\beta_w + \beta_{xw}x)^3 \theta_x}{\big((\beta_w + \beta_{xw}x)^2 \sigma^2 + \frac{\pi^2}{3}\big)^{3/2}}\Bigg\}
\end{gather*}
So, the standard error for the estimator of natural indirect effects is equal to 
\begin{equation*}
    se(\widehat{\log OR^{NIE}_{x,x^*}}) = \sqrt{\nabla f_1(\hat{\boldsymbol{\beta}}, \hat{\boldsymbol{\theta}}, \hat{\sigma}^2)^\top\, \Sigma \, \nabla f_1(\hat{\boldsymbol{\beta}}, \hat{\boldsymbol{\theta}}, \hat{\sigma}^2)},
\end{equation*}
where $\Sigma$ is defined in \eqref{sigma}.

%% file: tables.tex
\newpage

\begin{landscape}
\newgeometry{top = 13.5cm, bottom=0.1cm}
\section{Results of the simulation study}
\label{sec:res_tab}
\begin{table}[ht]
\centering
\tiny
\begin{tabular}{l r|rrrrr|rrrrr|rrrrr|rrrrr|rrrrr}
  \hline
\multicolumn{2}{c|}{} & \multicolumn{5}{c|}{$\beta_0 = -3$}  & \multicolumn{5}{c|}{$\beta_0 = -2$}  & \multicolumn{5}{c|}{$\beta_0 = -0.5$}  & \multicolumn{5}{c|}{$\beta_0 = 1$}  & \multicolumn{5}{c}{$\beta_0 = 2$} \\[5pt]
  & & TrueVal & Bias & Sd & Cv & CvB & TrueVal & Bias & Sd & Cv & CvB & TrueVal & Bias & Sd & Cv & CvB & TrueVal & Bias & Sd & Cv & CvB & TrueVal & Bias & Sd & Cv & CvB \\ 
  \hline
  $n$=150 & Our & 0.431 & -0.146 & 2.226 & 97.6  & 90.4  & 0.425 & -0.017 & 0.604 & 97.6  & 94.4  & 0.409 & -0.016 & 0.492 & 94.8  & 94.8  & 0.398 & 0.054 & 0.677 & 94.5  & 93.1  & 0.395 & 1.208 & 4.391 & 93.6  & 84.9  \\ 
  & VV &  & 0.124 & 3.910 &  & 95.1  &  & 0.017 & 0.582 &  & 96.5  &  & 0.061 & 0.455 &  & 95.7  &  & 0.274 & 0.817 &  & 90.5  &  & 2.445 & 6.858 &  & 80.7  \\ 
  & Gaynor &  & -0.562 & 8.855 &  & 91.2  &  & 0.046 & 0.679 &  & 94.2  &  & -0.014 & 0.490 &  & 94.8  &  & 0.108 & 0.771 &  & 92.8  &  & 3.563 & 12.503 &  & 83.7  \\ 
  & Exact &  & -0.152 & 2.246 &  & 90.3  &  & -0.011 & 0.597 &  & 94.5  &  & -0.015 & 0.482 &  & 94.8  &  & 0.045 & 0.683 &  & 92.8  &  & 1.158 & 4.302 &  & 84.9  \\ 
  & Nem &  & -0.337 & 2.914 & 93.8 &  &  & -0.048 & 0.626 & 96.3  &  &  & -0.025 & 0.486 & 94.8  &  &  & 0.072 & 0.693 & 93.0 &  &  & 1.477 & 5.045 & 83.5 &  \\ 
  \hline
  $n$=500 & Our & 0.431 & 0.016 & 0.457 & 97.1  & 93.9  & 0.425 & -0.019 & 0.336 & 95.0 & 93.8  & 0.409 & -0.007 & 0.270 & 94.3  & 94.9  & 0.398 & 0.029 & 0.356 & 94.1  & 95.1  & 0.395 & 0.037 & 0.543 & 94.4  & 92.3  \\ 
  & VV &  & 0.025 & 0.458 &  & 94.1  &  & 0.000 & 0.321 &  & 94.0 &  & 0.027 & 0.247 &  & 94.6  &  & 0.095 & 0.318 &  & 94.8  &  & 0.188 & 0.577 &  & 90.8  \\ 
  & Gaynor &  & 0.184 & 0.612 &  & 93.5  &  & 0.045 & 0.377 &  & 93.7  &  & -0.007 & 0.268 &  & 94.8  &  & 0.066 & 0.401 &  & 94.7  &  & 0.166 & 0.750 &  & 91.6  \\ 
  & Exact &  & 0.019 & 0.461 &  & 94.1  &  & -0.014 & 0.331 &  & 93.8  &  & -0.008 & 0.266 &  & 94.8  &  & 0.023 & 0.362 &  & 95.0 &  & 0.031 & 0.558 &  & 92.2  \\ 
  & Nem &  & -0.004 & 0.475 & 96.1  &  &  & -0.023 & 0.335 & 94.1  &  &  & -0.010 & 0.266 & 94.3  &  &  & 0.029 & 0.359 & 94.5  &  &  & 0.060 & 0.564 & 94.6  &  \\ 
  \hline
  $n$=1000 & Our & 0.431 & -0.009 & 0.326 & 96.3  & 92.9  & 0.425 & -0.012 & 0.233 & 95.3  & 94.0 & 0.409 & -0.004 & 0.178 & 95.6  & 94.6  & 0.398 & 0.005 & 0.244 & 93.6  & 94.8  & 0.395 & 0.011 & 0.372 & 91.8  & 93.2  \\ 
  & VV &  & 0.000 & 0.323 &  & 93.0 &  & 0.006 & 0.221 &  & 94.1  &  & 0.027 & 0.157 &  & 94.1  &  & 0.055 & 0.211 &  & 93.1  &  & 0.088 & 0.330 &  & 92.5  \\ 
  & Gaynor &  & 0.153 & 0.434 &  & 92.1  &  & 0.054 & 0.260 &  & 93.8  &  & -0.004 & 0.177 &  & 94.6  &  & 0.036 & 0.273 &  & 94.5  &  & 0.122 & 0.498 &  & 92.5  \\ 
  & Exact &  & -0.005 & 0.328 &  & 92.9  &  & -0.007 & 0.228 &  & 94.1  &  & -0.005 & 0.176 &  & 94.6  &  & 0.000 & 0.249 &  & 94.4  &  & 0.008 & 0.380 &  & 93.2  \\ 
  & Nem &  & -0.016 & 0.334 & 94.4  &  &  & -0.012 & 0.231 & 94.6  &  &  & -0.005 & 0.175 & 95.2  &  &  & 0.004 & 0.247 & 94.6  &  &  & 0.021 & 0.380 & 93.3  &  \\
  \hline
  $n$=5000 & Our & 0.431 & 0.001 & 0.142 & 96.6  & 94.9  & 0.425 & -0.004 & 0.102 & 95.6  & 95.0 & 0.409 & -0.001 & 0.081 & 94.3  & 94.5  & 0.398 & 0.011 & 0.108 & 93.8  & 94.1  & 0.395 & 0.014 & 0.154 & 94.6  & 94.5  \\ 
  & VV &  & 0.010 & 0.140 &  & 94.9  &  & 0.012 & 0.097 &  & 94.6  &  & 0.026 & 0.074 &  & 92.1  &  & 0.046 & 0.093 &  & 91.7  &  & 0.060 & 0.128 &  & 92.6  \\ 
  & Gaynor &  & 0.166 & 0.187 &  & 86.3  &  & 0.063 & 0.114 &  & 91.7  &  & -0.001 & 0.081 &  & 94.5  &  & 0.041 & 0.121 &  & 93.3  &  & 0.115 & 0.205 &  & 91.4  \\ 
  & Exact &  & 0.005 & 0.141 &  & 94.9  &  & 0.000 & 0.100 &  & 95.2  &  & -0.003 & 0.081 &  & 94.4  &  & 0.007 & 0.110 &  & 94.4  &  & 0.013 & 0.157 &  & 94.4  \\ 
  & Nem &  & 0.003 & 0.143 & 95.5  &  &  & -0.000 & 0.101 & 95.2  &  &  & -0.002 & 0.080 & 94.3  &  &  & 0.009 & 0.109 & 94.3  &  &  & 0.016 & 0.155 & 95.5  &  \\ 
  \hline
\end{tabular}
\caption{Natural Direct Effects:  bias, standard deviation and coverage of 95$\%$  CIs with delta/robust method and bootstrap}
\label{tab:nde}
\end{table}

\begin{table}[ht]
\centering
\tiny
\begin{tabular}{l r|rrrrr|rrrrr|rrrrr|rrrrr|rrrrr}
\hline
\multicolumn{2}{c|}{} & \multicolumn{5}{c|}{$\beta_0 = -3$}  & \multicolumn{5}{c|}{$\beta_0 = -2$}  & \multicolumn{5}{c|}{$\beta_0 = -0.5$}  & \multicolumn{5}{c|}{$\beta_0 = 1$}  & \multicolumn{5}{c}{$\beta_0 = 2$} \\[5pt]
  & & TrueVal & Bias & Sd & Cv & CvB & TrueVal & Bias & Sd & Cv & CvB & TrueVal & Bias & Sd & Cv & CvB & TrueVal & Bias & Sd & Cv & CvB & TrueVal & Bias & Sd & Cv & CvB \\ 
  \hline
  $n$=150 & Our & 0.322 & -0.005 & 0.547 & 84.2  & 88.9  & 0.319 & 0.015 & 0.389 & 88.8  & 92.3  & 0.317 & 0.022 & 0.348 & 88.6  & 94.4  & 0.320 & 0.014 & 0.455 & 87.1  & 93.1  & 0.323 & -0.043 & 0.688 & 79.6  & 78.8  \\ 
  & VV &  & -0.892 & 34.621 &  & 89.0 &  & 0.046 & 0.443 &  & 92.2  &  & 0.051 & 0.395 &  & 94.3  &  & 0.063 & 0.557 &  & 93.1  &  & -0.849 & 24.079 &  & 78.8  \\
  & Gaynor &  & 0.112 & 0.721 &  & 90.0 &  & 0.052 & 0.431 &  & 91.8  &  & 0.017 & 0.342 &  & 94.4  &  & 0.037 & 0.510 &  & 90.8 &  & 0.006 & 0.932 &  & 68.9 \\ 
  & Exact &  & 0.023 & 0.605 &  & 88.9  &  & 0.016 & 0.390 &  & 92.3  &  & 0.012 & 0.333 &  & 94.6  &  & 0.011 & 0.455 &  & 93.1  &  & -0.037 & 0.827 &  & 78.8  \\ 
  & Nem &  & 0.101 & 1.785 & 87.3 &  &  & 0.023 & 0.406 & 93.9  &  &  & 0.020 & 0.342 & 92.5  &  &  & 0.034 & 0.501 & 91.6  &  &  & -0.140 & 3.642 & 75.7 &  \\
  \hline
  $n$=500 & Our & 0.322 & -0.020 & 0.267 & 91.5  & 93.2  & 0.319 & 0.006 & 0.197 & 93.2  & 94.1  & 0.317 & 0.009 & 0.165 & 94.3  & 95.3  & 0.320 & -0.007 & 0.242 & 91.3  & 93.7  & 0.323 & 0.004 & 0.353 & 89.1  & 91.2  \\ 
  & VV &  & -0.006 & 0.288 &  & 93.0 &  & 0.018 & 0.211 &  & 94.0 &  & 0.019 & 0.176 &  & 95.0 &  & 0.007 & 0.262 &  & 93.6  &  & 0.032 & 0.405 &  & 90.9  \\ 
  & Gaynor &  & 0.069 & 0.349 &  & 92.2  &  & 0.038 & 0.219 &  & 93.3  &  & 0.007 & 0.162 &  & 95.2  &  & 0.029 & 0.267 &  & 92.7  &  & 0.095 & 0.465 &  & 89.4  \\ 
  & Exact &  & -0.014 & 0.276 &  & 93.2  &  & 0.006 & 0.198 &  & 94.0 &  & 0.004 & 0.159 &  & 95.1  &  & -0.007 & 0.241 &  & 93.7  &  & 0.013 & 0.369 &  & 91.0 \\ 
  & Nem &  & -0.012 & 0.283 & 93.5  &  &  & 0.008 & 0.200 & 95.1  &  &  & 0.006 & 0.161 & 96.0 &  &  & -0.003 & 0.247 & 93.9  &  &  & 0.025 & 0.397 & 91.2  &  \\
  \hline
  $n$=1000 & Our & 0.322 & -0.001 & 0.187 & 92.8  & 92.6  & 0.319 & 0.005 & 0.138 & 93.0 & 94.6  & 0.317 & 0.011 & 0.120 & 93.8  & 93.8  & 0.320 & 0.001 & 0.166 & 91.5  & 93.6  & 0.323 & -0.005 & 0.238 & 91.6  & 92.9  \\ 
  & VV &  & 0.010 & 0.200 &  & 92.6  &  & 0.014 & 0.146 &  & 94.1  &  & 0.018 & 0.127 &  & 94.0 &  & 0.010 & 0.176 &  & 93.6  &  & 0.009 & 0.256 &  & 93.1  \\ 
  & Gaynor &  & 0.091 & 0.245 &  & 92.1  &  & 0.036 & 0.152 &  & 94.5  &  & 0.008 & 0.118 &  & 93.6  &  & 0.039 & 0.181 &  & 92.9  &  & 0.093 & 0.310 &  & 92.1  \\ 
  & Exact &  & 0.004 & 0.193 &  & 92.6  &  & 0.005 & 0.138 &  & 94.5  &  & 0.006 & 0.116 &  & 93.7  &  & 0.001 & 0.165 &  & 93.6  &  & 0.001 & 0.245 &  & 93.0 \\ 
  & Nem &  & 0.005 & 0.196 & 95.0 &  &  & 0.006 & 0.139 & 94.6  &  &  & 0.008 & 0.117 & 95.0 &  &  & 0.003 & 0.167 & 93.9  &  &  & 0.004 & 0.250 & 93.7  &  \\ 
  \hline
  $n$=5000 & Our & 0.322 & -0.005 & 0.083 & 93.8  & 94.0 & 0.319 & -0.001 & 0.062 & 93.7  & 94.2  & 0.317 & 0.005 & 0.054 & 93.0 & 94.5  & 0.320 & -0.005 & 0.071 & 95.1  & 95.9 & 0.323 & -0.009 & 0.105 & 94.1  & 94.1  \\ 
  & VV &  & 0.001 & 0.087 &  & 93.6  &  & 0.004 & 0.065 &  & 94.9  &  & 0.011 & 0.056 &  & 94.1  &  & 0.000 & 0.074 &  & 96.0 &  & -0.002 & 0.111 &  & 93.5 \\ 
  & Gaynor &  & 0.082 & 0.107 &  & 86.6  &  & 0.027 & 0.069 &  & 92.2  &  & 0.003 & 0.053 &  & 94.2  &  & 0.035 & 0.077 &  & 91.6  &  & 0.096 & 0.135 &  & 86.6  \\ 
  & Exact &  & -0.002 & 0.084 &  & 93.9  &  & -0.002 & 0.062 &  & 94.2  &  & 0.002 & 0.052 &  & 94.1  &  & -0.005 & 0.071 &  & 95.9  &  & -0.005 & 0.108 &  & 93.8  \\ 
  & Nem &  & -0.002 & 0.085 & 94.6  &  &  & -0.001 & 0.062 & 94.8  &  &  & 0.002 & 0.053 & 94.4  &  &  & -0.005 & 0.071 & 96.2  &  &  & -0.005 & 0.108 & 94.2  &  \\ 
   \hline
\end{tabular}
\caption{Natural Indirect Effects: bias, standard deviation and coverage of 95$\%$  CIs with delta/robust method and bootstrap}
\label{tab:nie}
\end{table}
\end{landscape}